\documentclass[twocolumn]{aastex631}

\usepackage{savesym}
\usepackage{graphicx}	
\usepackage{amsmath}	
\usepackage{amssymb}	

\renewcommand{\figurename}{Fig.}

\savesymbol{longtable}
\usepackage{xcolor}

\usepackage[utf8]{inputenc}
\usepackage[english]{babel}

\restoresymbol{SUB}{longtable}

\usepackage{physics}

\begin{document}

\title{Marginal Role of the Electrostatic Instability in the GeV-scale Cascade Flux from 1ES 0229+200}

\author[0009-0008-0835-2795]{Mahmoud Alawashra}
\affiliation{Deutsches Elektronen-Synchrotron DESY, Platanenallee 6, 15738 Zeuthen, Germany}
\email{mahmoud.al-awashra@desy.de}

\author[0000-0003-3444-3830]{Ievgen Vovk}
\affiliation{Institute for Cosmic Ray Research, The University of Tokyo, 5-1-5 Kashiwa-no-Ha, Kashiwa City, Chiba, 277-8582, Japan}

\author[0000-0001-7861-1707]{Martin Pohl}
\affiliation{Deutsches Elektronen-Synchrotron DESY, Platanenallee 6, 15738 Zeuthen, Germany}
\affiliation{Institute for Physics and Astronomy, University of Potsdam, D-14476 Potsdam, Germany}

\received{30-Oct-2024}
\revised{26-Nov-2024}
\accepted{29-Nov-2024}
\published{-}
\submitjournal{The Astrophysical Journal}

\begin{abstract}

Relativistic pair beams produced in the intergalactic medium (IGM) by TeV gamma rays from blazars are expected to generate a detectable GeV-scale electromagnetic cascade, yet this cascade is absent in the observed spectra of hard-spectrum TeV emitting blazars. This suppression is often attributed to weak intergalactic magnetic fields (IGMF) deflecting electron-positron pairs out of the line of sight. Alternatively, it has been proposed that beam-plasma instabilities could drain the energy of the beam before they produce the secondary cascades. Recent studies suggest that the modification of beam distribution due to these instabilities is primarily driven by particle scattering, rather than energy loss. In this paper, we quantitatively assess, for the blazar 1ES 0229+200, the arrival time of secondary gamma rays at Earth from the beam scattering by the electrostatic instability. We first computed the production rates of electron-positron pairs at various distances using the Monte Carlo simulation CRPropa. We then simulated the feedback of the plasma instability on the beam, incorporating production rates and inverse-Compton cooling, to determine the steady-state distribution function. Our findings reveal that the time delay of the GeV secondary cascade arrival due to instability broadening is on the order of a few months. This delay is insufficient to account for the missing cascade emission in blazar spectra, suggesting that plasma instabilities do not significantly affect IGMF constraints.

\end{abstract}




\section{Introduction}

Blazars are active galactic nuclei, characterised by the orientation of their relativistic jet towards the Earth. Observations conducted with the Fermi-LAT and the imaging atmospheric Cerenkov telescopes (such as VERITAS, MAGIC, and HESS) have revealed the presence of bright GeV-TeV gamma-ray emission from several blazars \citep{1752,2010A&A...520A..83H}. The propagation of those very-high-energy gamma rays over cosmological distances has been a topic of extensive theoretical and observational investigations, as it offers a unique probe into the intergalactic medium (IGM) \citep{galaxies10020039}.

TeV gamma rays interact with the extragalactic background light (EBL), producing focused beams of electron-positron pairs. These pairs are expected to lose their energy through inverse Compton scattering on the cosmic microwave background (CMB) photons, generating an electromagnetic cascade in the GeV energy range \citep{1967PhRv..155.1408G,1970RvMP...42..237B}. This GeV cascade is detectable for blazars with hard intrinsic spectrum that peak well beyond the TeV gamma-ray energies \citep{2009PhRvD..80l3012N}. However, the GeV secondary cascade has not yet been detected in the gamma-ray spectra of those hard-spectrum TeV emitting blazars \citep{2010Sci...328...73N,Acciari_2023,Aharonian_2023}. 

The absence of GeV cascade emission in the gamma-ray spectra of blazars can be attributed to the deflection of TeV pairs by intergalactic magnetic fields (IGMF) \citep{PhysRevD.80.023010,PhysRevD.80.123012,2010Sci...328...73N,2011A&A...529A.144T,Takahashi_2011,Vovk_2012,Durrer_2013}. This deflection leads to an extended emission or/and a time delay of the cascade emission. As a result, the observed blazar spectra are used to set lower limits on the strength of the IGMF.

Those IGMF limits are based on the assumption that the inverse Compton cooling dominates the evolution of the induced pair beams. However, this assumption neglects the effect of the plasma instabilities that arise from collective interactions between the relativistic pair beams and the denser IGM background plasma. Early studies viewed the plasma instabilities as an efficient cooling mechanism that is faster than the inverse Compton cooling \citep{Broderick_2012,Schlickeiser_2012,Miniati_2013,Schlickeiser_2013,2014ApJ...790..137B,Sironi_2014,Chang_2014,Supsar_2014,2016ApJ...833..118C,2016A&A...585A.132K,2017A&A...607A.112R,Vafin_2018,Vafin_2019,AlvesBatista:2019ipr,shalaby_broderick_chang_pfrommer_puchwein_lamberts_2020}.

Later studies have shown that the efficiency of the instability energy loss is limited by several factors in the IGM. The potential weak IGMF could quench the instability by increasing the beam opening angle \citep{Alawashra_2022}. The instability feedback on the beam results in the scattering of the beam particles in the transverse direction suppressing the instability as well \citep{Perry_2021,Alawashra_2024}.

Whether the scattering by the instability feedback could significantly impact the secondary cascade arrival time at Earth remains unanswered. The recent studies of such feedback include \citet{Perry_2021}, who used a simplified 1D beam profile, and \citet{Alawashra_2024}, who used a realistic 2D pair distribution generated by the annihilation of high-energy gamma rays with the extragalactic background light at a distance of 50 Mpc from a fiducial BL Lac source \citep{Vafin_2018}. In order to accurately assess the impact of instability on the observed GeV cascade flux, one needs to calculate the beam scattering at various distances in the IGM using a specific BL Lac source.

In this study, we use the blazer 1ES 0229+200~\citep[responsible for the strongest single-source IGMF constraints thus far,][]{Acciari_2023, Aharonian_2023} to study the impact of the instability on GeV-band cascade emission. We focused solely on diffusion caused by instability. We compare the resulting time delay due to the instability scattering with that of one of the magnetic deflections at the end. 

We compute the pair creation rate at several distances in the IGM from the source using the Monte Carlo simulation CRpropa. Then we solve numerically the time evolution of the pair beam distribution including diffusive feedback and inverse Compton cooling on the CMB photons. We aim to compute the cascade arrival time delay due to the broadened beam profiles at the steady state between diffusion, inverse Compton cooling, and pair creation.

MeV cosmic-ray electrons in the IGM could Landau damp the plasma oscillations with oblique angles to the beam axis \citep{Yang_2024}. Throughout our calculations, we omitted any possible effect of this damping and the IGMF as well. The linear Landau damping might limit the growth of the instability at the oblique angles and allow for efficient growth of the parallel modes. A qualitative investigation of linear Landau damping, combined with the nonlinear evolution of the instability, is necessary to fully assess its impact and it's left for future studies.

The structure of this paper is as follows. In section \ref{sec:2}, we calculate the pair creation rate for a 1ES 0229+200-like source. In section \ref{sec:3}, we present the beam transport equation under the quasilinear theory of the beam-plasma system including Fokker-Planck diffusion, inverse Compton cooling and beam creation rates. We present the numerical simulation setup for the beam evolution calculations in section \ref{sec:4} and the results in section \ref{sec:5}. Finally, we conclude in section \ref{sec:con}.

\section{1ES 0229+200 induced pair beams creation rates}\label{sec:2}

Before we study the instability feedback in the evolution of the blazer-induced pair beam, it is necessary to evaluate the production rates of the pair beam at different distances in the IGM. For this purpose, we consider the BL Lac source 1ES 0229+200 which has an intrinsic gamma-ray emission spectrum that is harder than $E^{-2}$ extends to a few TeV. The advantage of this spectrum is that the cascade emission should be predominantly in the GeV energy band which the Fermi-LAT operates at \citep{Abdo_2010}.

We used the Monte Carlo code, CRpropa \citep{Batista_2016,AlvesBatista_2022} to find the energy dependence of those creation rates at several distances from the source. For a gamma-ray source like 1ES 0229+200, we used the following power-law intrinsic spectrum
\begin{equation}
F(E_{\gamma}, z = 0) = \, N_0 \left( \frac{E_{\gamma}}{\text{GeV}} \right)^{-\alpha} \exp \left( -\frac{E_{\gamma}}{E_{\gamma,\text{max}}} \right),
\end{equation}
with normalization $N_0 = 2.6 \times 10^{-10}~\text{ph} / (\text{cm}^2 \, \text{s} \, \text{GeV})$, power index  $\alpha = 1.7$ and cut-off energy $E_{\gamma, \text{max}} = 10 $~TeV. The source was located at a redshift of 0.144 and the EBL model used was taken from \citet{Franceschini_2008}. We used the following cosmological parameters through this paper, $H_0 = 69.6$ km s$^{-1}$ Mpc$^{-1}$, $\Omega_m = 0.286$ and $\Omega_\Lambda = 0.714$.

\begin{figure}
\centering
    \includegraphics[width=\columnwidth]{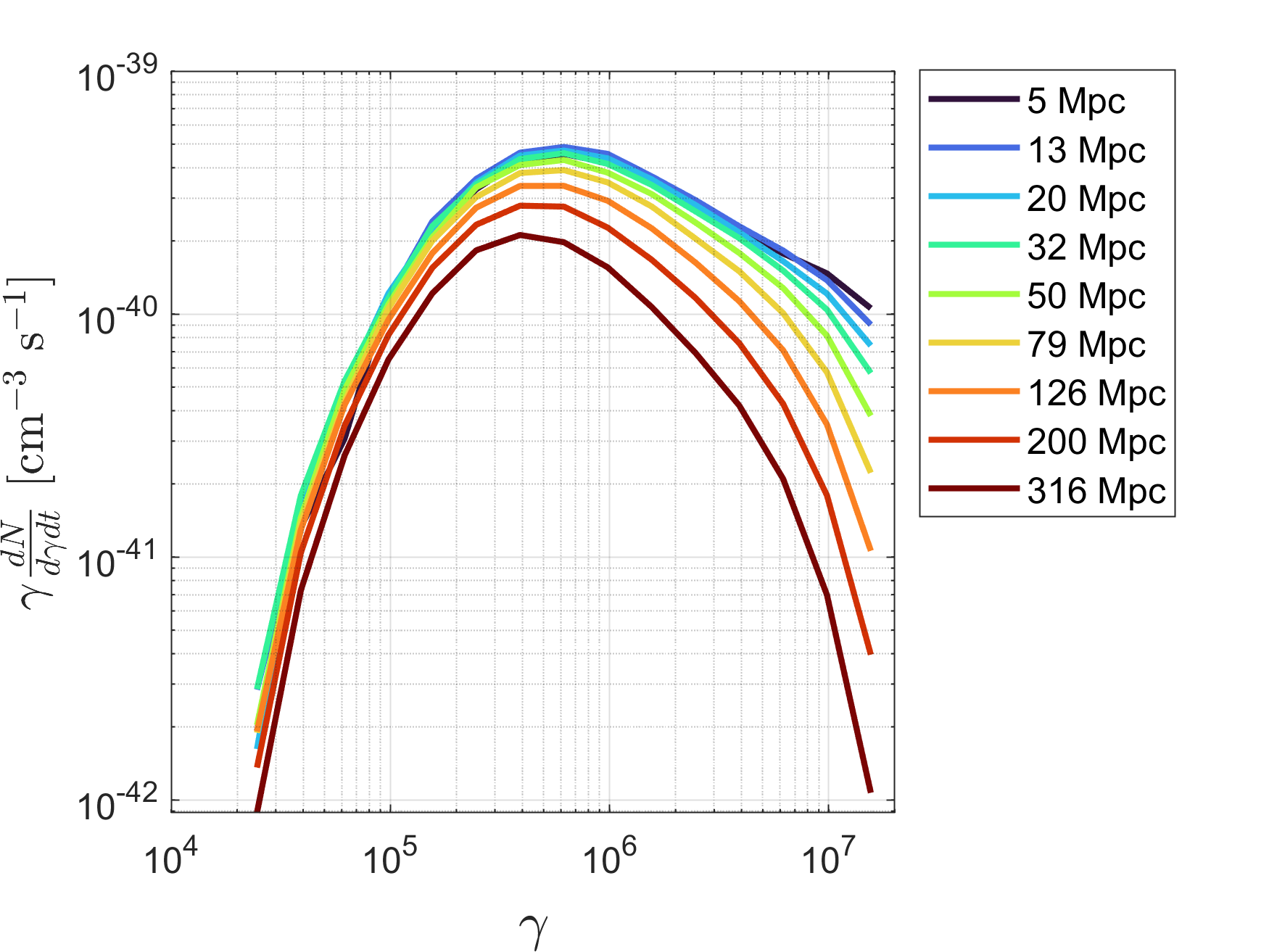}
    \caption{Pair beams creation rates, $\gamma \frac{df_\gamma}{dt}\equiv \gamma\frac{dN}{d\gamma dt}$, from CRpropa at different distances from 1ES 0229+200 like source. All the rates are normalized to the Earth luminosity distance from the source of 687 Mpc.}
    \label{fig:Rates}
\end{figure}

The differential pair creation rates, $\frac{df_\gamma}{dt} \equiv \frac{dN}{d\gamma dt}$, are shown in Fig.\ref{fig:Rates} for various distances from the source. All the curves in Fig.\ref{fig:Rates} are normalized to the Earth-source luminosity distance, which is 687 Mpc. In addition to the implicit $R^{-2}$ distance dependence in Fig.\ref{fig:Rates}, we observe a decline in the production of pairs with the highest energies as the distance in the IGM increases. This decline is attributed to the EBL's absorption of high-energy gamma rays.

To use those rates in the beam continuity equation we need to transform Lorentz factor dependence to momentum dependence, $(p,\theta, \varphi)$, in spherical coordinates, where the beam propagates along the polar axis. The angular dependence can be expressed as follows
\begin{equation}\label{eq:1}
    Q_{ee}(p,\theta) = \frac{df}{dp^3dt} = \frac{1}{ 2 \pi m_ec\beta p^2}\frac{df_\gamma}{dt}f_{\cos{\theta}}(\gamma,\theta),
\end{equation}
where $f_{\cos{\theta}}(\gamma,\theta) \equiv \frac{dN}{\theta d \theta}$ is the angular differential part and $\beta$ is the normalized speed. The angular distribution of the produced pairs by the gamma rays follows a narrow band around the angles, $\theta \sim \gamma^{-1}$, with the angular differential part of the distribution given approximately by \citep{Broderick_2012,Miniati_2013,Vafin_2018}
\begin{equation}\label{eq:2}
    f_{\cos{\theta}}(\gamma,\theta) =\frac{2}{ \Delta\theta^2}\exp{-\big(\frac{\theta}{\Delta\theta}\big)^2},
\end{equation}
where the angular spread of the newly produced particles is given by $\Delta\theta = \frac{1}{\gamma}$. We will use those rates in the numerical calculations of the Fokker-Planck diffusion of the instability feedback in section \ref{sec:4}.

\section{Pair Beam temporal evolution}\label{sec:3}

In this section, we introduce the theoretical model governing the temporal evolution of the pair beam within the IGM. The key processes influencing this evolution include beam diffusion driven by plasma instability feedback, pair production, and inverse Compton cooling. These processes are encapsulated in the following transport equation
\begin{equation}\label{eq:f}
\begin{split}
     \frac{\partial f(p,\theta)}{\partial t} = & \frac{1}{p^2}\frac{\partial}{\partial p}\left(- \Dot{p}_{\text{IC}} p^2 f(p,\theta)\right) + Q_{ee}(p,\theta)  \\ & + \frac{1}{p^2\theta}\frac{\partial}{\partial \theta}\left(\theta D_{\theta\theta}(\theta,t) \frac{\partial f(p,\theta)}{\partial \theta}\right).
\end{split}
\end{equation}

The first term on the right-hand side represents the momentum advection caused by inverse Compton scattering on cosmic microwave background (CMB) photons. Since the Lorentz factors of the produced pairs are within the Thomson scattering regime, where $(4\epsilon_{\text{CMB}} \gamma \ll m_e c^2)$, the cooling rate is expressed as
\begin{equation}
    \Dot{p}_{\text{IC}} = - \frac{4}{3} \sigma_\text{T} u_\text{CMB} \gamma^2,
\end{equation}
where $\sigma_\text{T}$ is the Thompson cross-section, and $u_\text{CMB}$ is the CMB energy density, which scales with redshift as $(1+z)^4$ \citep{Broderick_2012}. We used a CMB energy density of 0.26 eV/ cm$^{-3}$ at redshift $z=0$ \citep{Fixsen_2009}. The angular advection caused by inverse Compton scattering is negligible in the Thomson regime, as the total angular deflection it induces is smaller than the characteristic production angles,  $\theta \sim \gamma^{-1}$. 

The second term is the source term for pair creation resulting from gamma-ray annihilation with the EBL photons, as detailed in section \ref{sec:2} for the source 1ES 0229+200. The third term captures the feedback of electrostatic unstable waves on the beam, modelled through Fokker-Planck diffusion \citep{Brejzman_1974}. This quasilinear approach of the instability feedback is valid when the wave energy density is significantly lower than that of the IGM plasma. We have neglected the momentum diffusion of the instability feedback as the $\theta \theta$ diffusion dominates the feedback of the instability \citep{Perry_2021,Alawashra_2024}.

The diffusion coefficient, $D_{\theta \theta}$, is defined by the following resonance integral over the oscillating electric field energy density, $W$, \citep{1971JETP...32.1134R}
\begin{equation}\label{eq:D}
    D_{\theta \theta}(\boldsymbol{p},t) = \pi e^2 \int d^3\boldsymbol{k} W(\boldsymbol{k},t) \frac{k_\theta k_\theta}{k^2} \delta(\boldsymbol{k}\cdot\boldsymbol{v}-\omega_p),
\end{equation}
where the electric charge, $e$, is given in cgs units. The wave vector, $\boldsymbol{k}$, is given in spherical coordinates, $(k,\theta',\varphi')$, where the beam axis is the z-direction. $k_\theta$ is the projection of the wave vector to the angler spacial component ($\theta$-direction), $k_\theta = \boldsymbol{k}\cdot\boldsymbol{\hat{\theta}} = k[\sin{\theta'}\cos{\theta}\cos{\varphi'}-\cos{\theta'}\sin{\theta}]$.

Due to the azimuthal symmetry of the beam and without losing the generality, we set $\varphi =0 $ and integrate over $\varphi'$, yielding

\begin{equation}\label{eq:Dtt}
\begin{split}
        D_{\theta \theta}  = &  \frac{\pi m_e \omega_p^2}{n_e} \int_{\omega_p/c}^\infty k^2dk \int_{\cos{\theta_1'}}^{\cos{\theta_2'}} d\cos{\theta'} \frac{W(\boldsymbol{k}) }{kv_b} \\ &  \times \frac{\xi^2 }{\sqrt{(\cos{\theta'}-\cos{\theta_1'})(\cos{\theta_2'}-\cos{\theta'})}},
\end{split}
\end{equation}
where
\begin{equation}
    \xi = \frac{\cos{\theta}\frac{\omega_p}{kv_b}-\cos{\theta'}}{\sin{\theta}},
\end{equation}
and $v_b \approx c (1-\frac{1}{2\gamma^2})$ is the particle speed for Lorentz factor $\gamma$. The boundaries of the $\cos{\theta'}$ integration are fixed by the resonance condition to
\begin{equation}
    \cos{\theta_{1,2}'} = \frac{\omega_p}{kv_b}\left[\cos{\theta}\pm\sin{\theta}\sqrt{\left(\frac{kv_b}{\omega_p}\right)^2-1}\right].
\end{equation}
For a robust numerical integration over the electric field energy density, $W$, we do a change of variables to the coordinates $(k_\perp,\theta^{R})$ yielding equation A11 in \citet{Alawashra_2024}. Where $\theta^{R} = \left(\frac{ck_{||}}{\omega_p}-1\right)/(ck_\perp/\omega_p)$, with $k_\perp$ and $k_{||}$ being the perpendicular and parallel components of the wave vector to the beam axis respectively. We also ignore the dependency of the diffusion coefficients on the Lorentz factor, $\gamma$, as it is negligible compared to the beam angle, $\theta$.

The evolution of the electric field fluctuations, $W$, is governed by the following equation
\begin{equation}\label{eq:W}
    \frac{\partial W (\vb{k},t)}{\partial t} = 2 (\omega_i (\vb{k},t) + \omega_c(k)) W(\vb{k}), 
\end{equation}
where $\omega_i(\vb{k})$ is the linear growth rate of the electrostatic instability that is given by \citep{1990}
\begin{equation}\label{eq:wi}
    \omega_i(\boldsymbol{k},t) = \omega_p \frac{2\pi^2 e^2}{k^2} \int d^3\boldsymbol{p} \left(\boldsymbol{k}\cdot\frac{\partial f(\boldsymbol{p},t)}{\partial \boldsymbol{p}}\right) \delta(\omega_p - \boldsymbol{k}\cdot\boldsymbol{v}_b),
\end{equation}
and $\omega_c$ is the collisional damping rate \citep{Tigik_2019}
\begin{equation}\label{eq:damp}
     \omega_c(k) = - \omega_p \frac{g}{6\pi^{3/2}}\frac{1}{(1+3k^2\lambda_D^2)^3}.
\end{equation}
Here $\omega_p = (4\pi n_e e^2/m_e)^{1/2}$ is the plasma frequency of the intergalactic background plasma with density $n_e$, $g=(n_e\lambda_D^3)^{-1}$ is the plasma parameter, $\lambda_D = 6.9 \text{ cm} \sqrt{\frac{T_e/ K}{n_e/\text{cm}^{-3}}} $ is the Debye length, and $T_e = 10^{4} K$ is the electrons temperature in the IGM.

The system of the equations \ref{eq:f} and \ref{eq:W} describe the nonlinear evolution of the beam-plasma system along with the production and cooling of the beam. The plasma waves influence the beam particles through the diffusion, $D_{\theta\theta}$ (equation \ref{eq:D}), and the beam shapes the growth of the waves by the linear growth rate, $\omega_i$ (equation \ref{eq:wi}). To determine the angular dispersion of the pair beam scattered by instability feedback, we need to solve those coupled temporal equations governing the evolution of both the beam and the waves (equations \ref{eq:f} and \ref{eq:W}). 

The numerical solution of equations \ref{eq:f} and \ref{eq:W} requires time steps significantly shorter than the inverse Compton cooling times of the pairs, due to the large linear growth rates involved \citep{Alawashra_2024}. However, the pair beams themselves propagate over cosmological distances, spanning billions of years. To address this challenge of scale separation, we incorporate beam production rates at various distances within the IGM, derived from CRpropa Monte Carlo simulation. These production rates are then utilized in the beam-plasma calculations to determine the resulting angular spread of the beams once the system reaches a steady state. In the next sections, we apply this method to the pair beams induced by the gamma rays from the source 1ES 0229+200.

\section{Numerical setup}\label{sec:4}

\begin{figure}
\centering
        \includegraphics[width=\columnwidth]{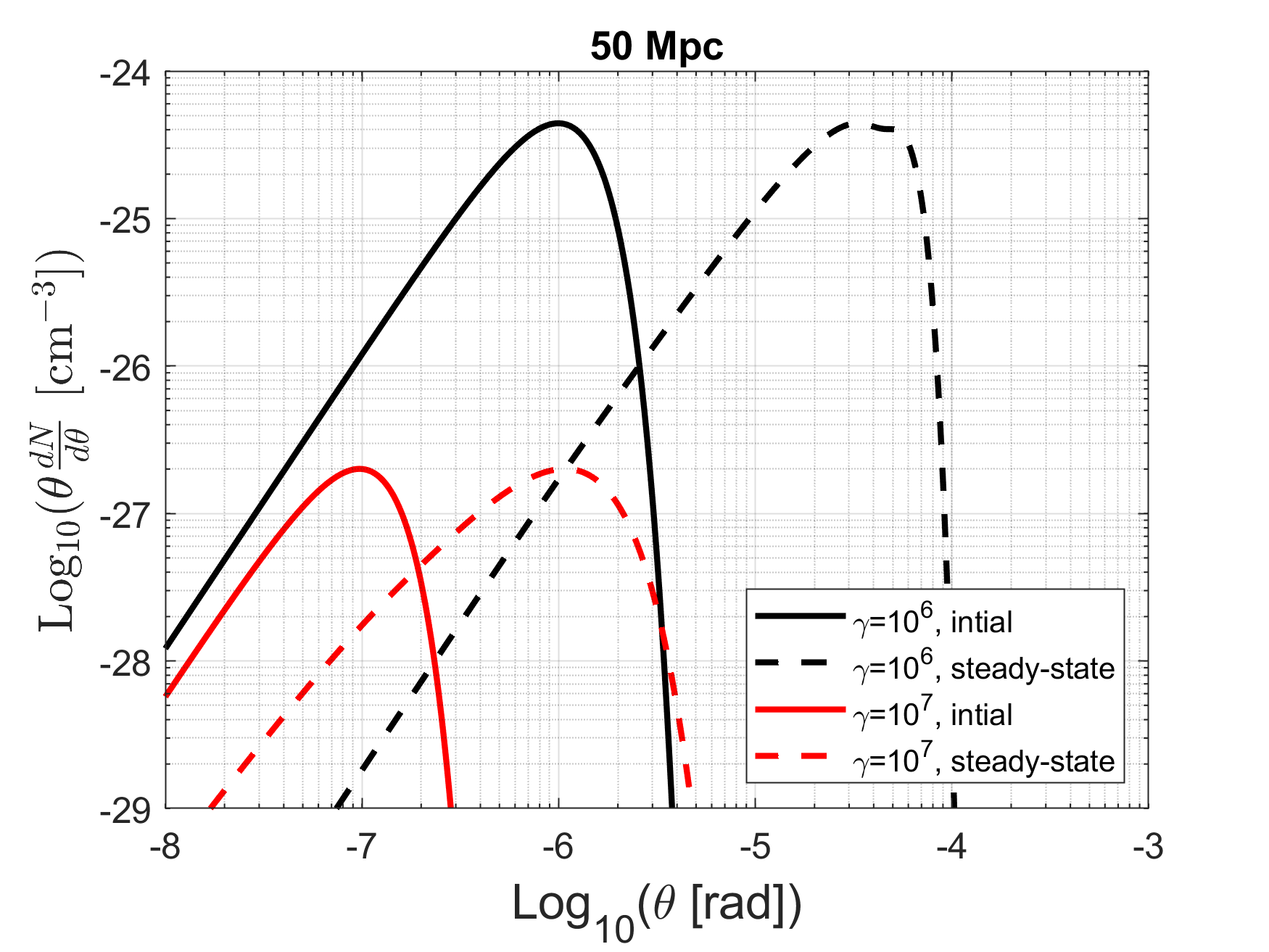}
        \caption{The beam angular profiles for given Lorentz factors ($\gamma = 10^6, 10^7$) before (initial) and after (steady-state) the diffusion by the instability. The calculations in this figure are performed for the pair beams produced at a distance of 50 Mpc from the source.}
        \label{fig:Ng6g7}
\end{figure}

\begin{figure}
\centering
        \includegraphics[width=\columnwidth]{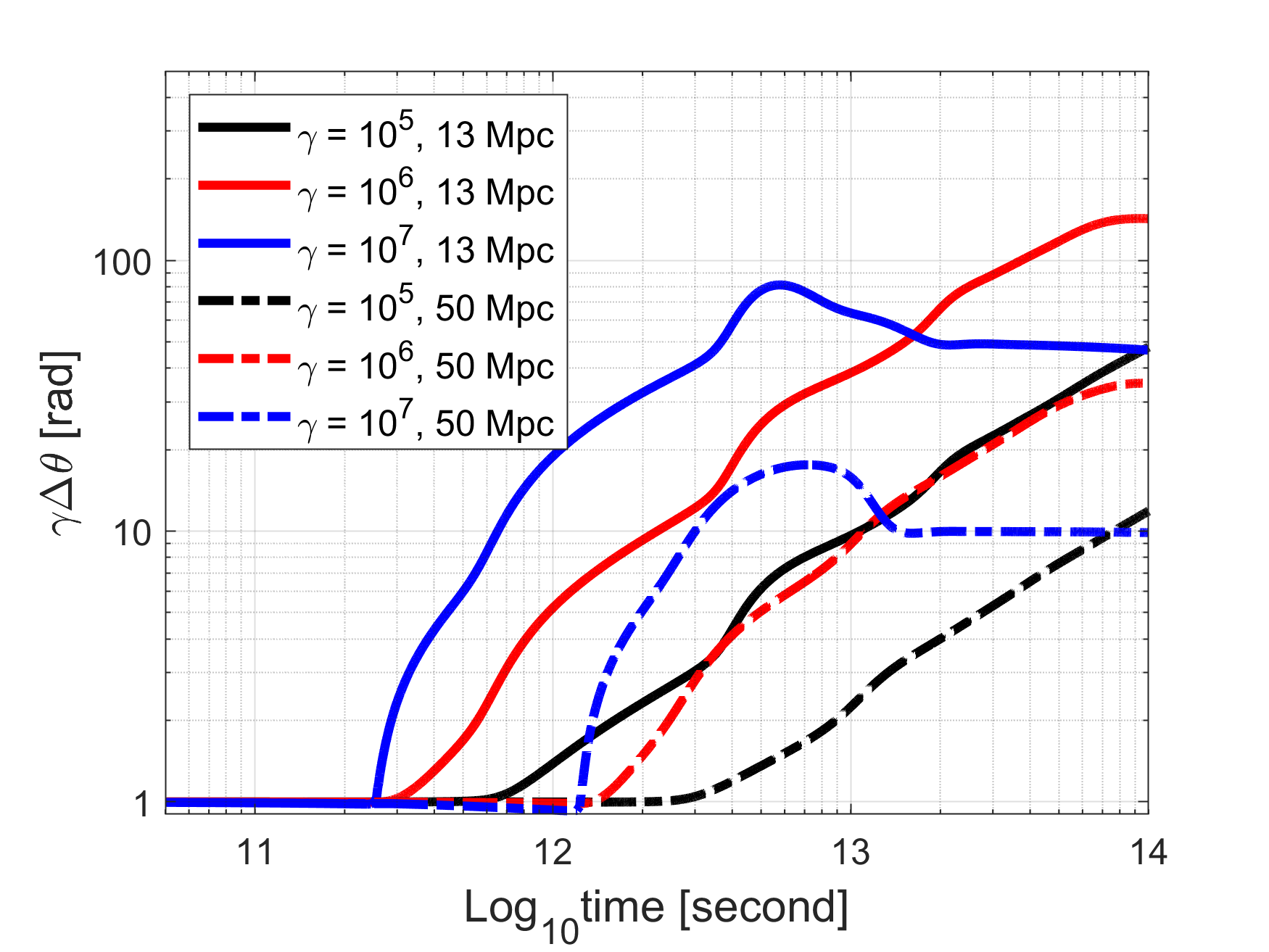}
        \caption{The time evolution of the beam angular spread for selected Lorentz factors at the distances of 13 Mpc and 50 Mpc from the source 1ES 0229+200.}
        \label{fig:deltatheta}
\end{figure}

We first calculated by the CRpropa Monte Carlo code the beam production rates, $Q_{ee}$, at logarithmically binned distances from a 1ES 0229+200-like source. After that for each distance, we numerically solve the beam transport equation (\ref{eq:f}) including inverse Compton cooling, pair production, and the angular diffusion. We set the initial condition to an accumulated beam over a time step of a year, shorter than the initial instability growth time ($\omega^{-1}_{i,\text{max}}\sim 1 - 100$ yr) and the expected inverse Compton cooling time ($\tau_{\text{IC}}\sim 2500$ yr$\left(\frac{10^6}{\gamma}\right)$). We employ operating split in equation \ref{eq:f}, where we solve the inverse Compton part using the Crank-Nicolson scheme with forward momentum derivative \citep{10.1046/j.1365-8711.1999.02538.x}, and we solve the $D_{\theta \theta}$ diffusion using the Crank-Nicolson scheme with central angular derivative \citep{Alawashra2024}. 

The beam transport equation (\ref{eq:f}) is integrated together with the waves' spectral evolution equation \ref{eq:W}. Where the wave spectrum determines the diffusion coefficients (equation (\ref{eq:D})) and the beam distribution shapes the linear growth rate of the instability (equation (\ref{eq:wi})). We solve the wave equation (\ref{eq:W}) using the forward time-centred space scheme after transforming the left-hand side to the time derivative of the natural logarithm of the electric field energy density, $W$. The initial wave energy density corresponds to the IGM background thermal fluctuation level \citep{Vafin_2019}.

We used a dynamic time step of the shortest one between the $\omega_{i,\text{max}}^{-1}$, $\omega_c^{-1}$, and the inverse of the fastest rate of change of the beam distribution. We also imposed a maximal time step of $10^{11}$ seconds. We tested this with 10 times smaller time steps. We employed the same coordinates used by \citet{Alawashra_2024} to resolve the narrow wave spectrum, $(ck_\perp/\omega_p,\theta^R)$ where $\theta^R = (ck_{||}/\omega_p-1)/(ck_\perp/\omega_p)$. $k_\perp$ and $k_{||}$ are the perpendicular and the parallel wave-vector components to the beam propagation direction respectively. 

\section{Results and discussion} \label{sec:5}

Initially, the instability grows due to the production of focused pairs with an angular spread of $\gamma^{-1}$. Once the fluctuating electric field is strong enough, it scatters the pairs to larger angles. When the pairs with certain Lorentz factors reach their IC cooling time, the beam angular profile at those Lorentz factors reaches a confined steady state where the beam opening angle is higher than the initial $\gamma^{-1}$ by some factor. This confined steady state is pronounced for the pairs with the highest Lorentz factors as their IC cooling times are short. In Fig.\ref{fig:Ng6g7} we show the confined steady state of the beam after the expansion by the instability diffusion and compare it with the initial profile at the distance of 50.1 Mpc from the source. For the pairs whose IC cooling time has not been reached yet, the angular profile expands continuously under a quasi-steady state.

In Fig.\ref{fig:deltatheta} we show the time evolution of the beam expansion for selected Lorentz factors at the corresponding distances of 12.6 Mpc and 50.1 Mpc from the source. We see that the expansion is higher for closer distances to the source as the instability growth rate scales with pair density that is higher closer to the blazer. We stopped the simulations at around $10^{14}$ seconds as we noted that pairs with Lorentz factors $10^6$ reach the confined steady state. Those pairs are the ones responsible for producing the GeV-scale cascade emission. 

The numerical simulations indicate a quasi-steady state in the electric-field amplitude, $W$, where the linear growth rate and the collisional damping rate balance for resonant wave numbers. However, due to the expansion of the beam to larger angles, the instability extends to higher values of the parallel wave number.

The original blazar gamma-ray beam has an opening angle of a few degrees, and further out in the IGM it locally is well collimated, as the gamma-ray emitter is essentially a point source. Therefore the widening of the pair beam due to the instability does not affect the width of the gamma-ray beam and its arrival flux. The only relevant effect of this instability widening of the beam is the arrival time delay of the Gev secondary cascade ar Earth.

For each distance in the IGM from the blazer, we use the resulting beam angular spread, $\Delta \theta (\gamma)$, for each Lorentz factor to calculate the arrival time delay of the cascade emitted at the corresponding distance in the IGM. The time delay of secondary gamma rays emitted by pairs that have undergone a deflection by an angle of $\Delta \theta$ from the primary gamma-ray propagation direction is given by the following formula \citep{2009PhRvD..80l3012N}
\begin{equation}
    \Delta t_{\text{delay}} \simeq (1+z) \frac{D_c \Delta \theta^2}{2c} \left(1-\frac{D_c}{D_b}\right),
\end{equation}
where $D_c$ is the distance between the emitting pairs and the blazer, $D_b$ and $z$ are the angular diameter distance and the redshift of the blazar. For a 1ES 0229+200-like source that we employed in our simulation setup, the angular diameter distance from the earth is $D_b=524.9$ Mpc.

\begin{figure}
\centering
        \includegraphics[width=\columnwidth]{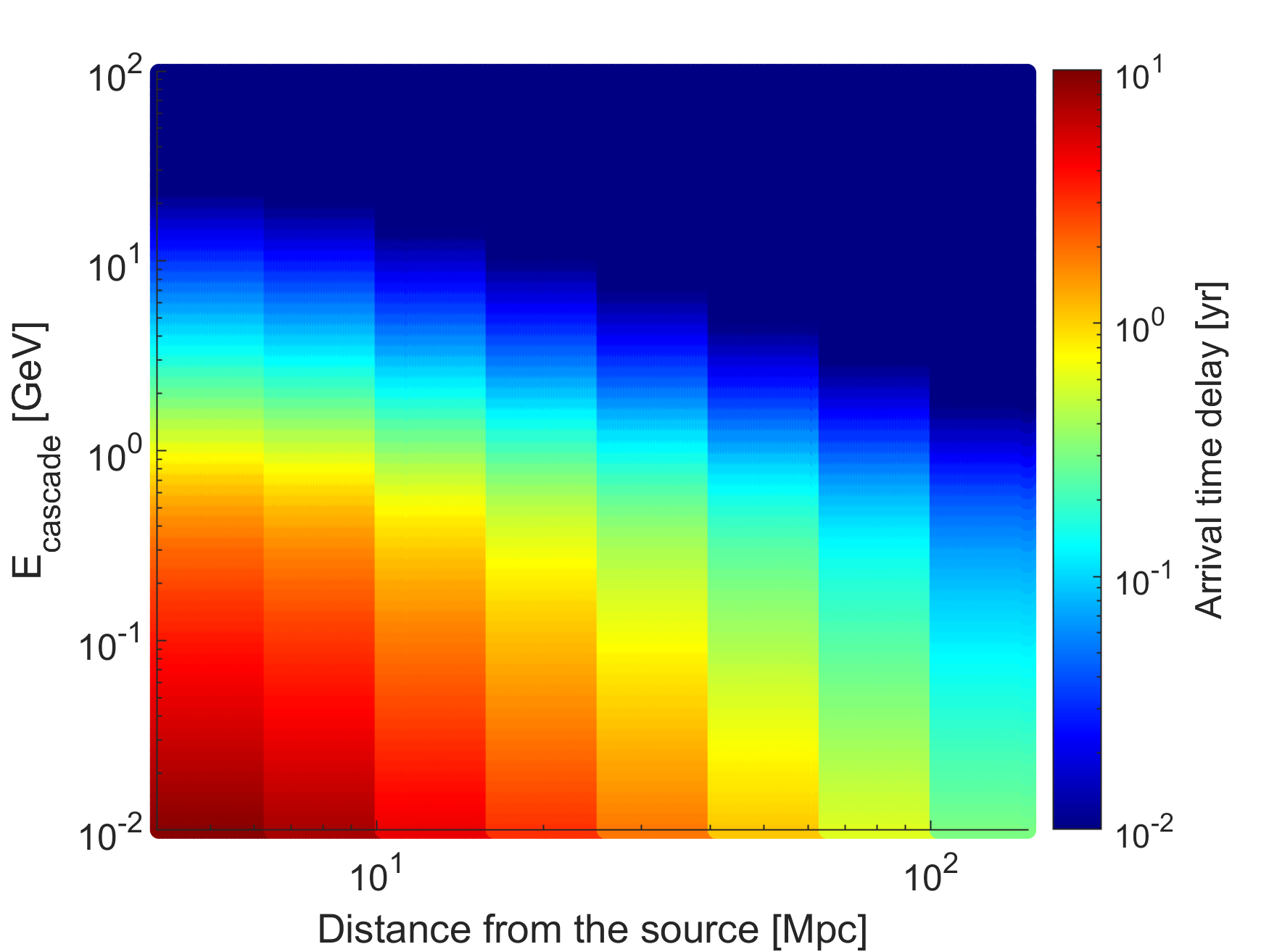}
        \caption{The arrival time delay at Earth of the cascade emission for a 1ES 0229+200-like source due to the scattering of the pair beams by the instability. The time delay is given for a cascade produced at a certain distance from the source (x-axis) and with a given energy (y-axis).}
        \label{fig:timeDelay}
\end{figure}

In Fig.\ref{fig:timeDelay}, we show the arrival time delay of the cascade emission for a 1ES 0229+200-like source due to the scattering of the pair beams by the instability. The time delay is given for the cascade photons emitted at certain distance from the source and with energy, $E_{\text{cascades}}$. Where we approximated the cascade photons energy produced from a beam particle with Lorentz factor,$\gamma$, as $E_{\text{cascades}}= \frac{4}{3} \epsilon_\text{CMB} \gamma^2$, where the average CMB photon energy, $\epsilon_\text{CMB}$, is $630\mu$eV \citep{Fixsen_2009}. This approximation provides us with the maximum time delay of the cascade photons at certain energy. For an accurate estimate one will need to deal with the full distribution of the secondary gamma rays produced by a beam particle. 

We see in Fig.\ref{fig:timeDelay}, that the time delay is the largest for the cascade photons, with the lowest energies emitted closer to the blazer. The delay decreases with the distance as the pair density drops with the distance and so does the instability growth. Cascade photons of lower energy are produced by the pairs with smaller Lorentz factors and therefore those pairs are produced with higher opening angles, $\Delta \theta \sim \gamma^{-1}$. This yields a larger time delay with the cascade photon energy decreasing, as the corresponding emitting pairs start diffusing from larger angles. 

We also examined the time delay for the cases of flaring activity of the source and other sources with higher and lower luminosities. We observed an increment by a factor of $\sqrt{C}$ in the beam angular spread when the source luminosity was inflated by a factor $C$. The reason for this is the balance between the linear growth rate and the collisional damping rate at the saturation. Since the linear growth rate scales proportionally to the beam density, $n_b$, and inversely with the square of the angular spread, $\Delta\theta$, $\omega_i \propto \frac{n_b}{\Delta\theta^2}$, the increment in the density by a factor $C$ requires an increment by a factor of $\sqrt{C}$ in the angular spread to keep the balance between the growth rate and the collisional damping rate at instability saturation. The increment of the angular spread by a factor $\sqrt{C}$ results in the increment of the cascade time delay by the factor $C$.

Therefore, even in the case of flaring activities of the blazar where the luminosity is higher by a factor of three comparable to the found TeV-band variability amplitude of 1ES~0229+200 \citep{Acciari_2023}, the arrival time delay of the bulk of the GeV-band cascade is still in the order of several months. The time delay of the secondary cascade due to the instability diffusion is subdominant compared to the time delay due to beam deflection by femto Gauss IGMF. Therefore, we conclude that the missing GeV-band cascade is more likely attributed to the existence of the IGMF rather than the plasma instability.

\section{Conclusions}\label{sec:con}

In this study, we investigated the broadening of the blazar-induced pair beams due to the diffusive feedback of the electrostatic beam-plasma instability in the IGM. By employing a realistic gamma-ray intrinsic spectrum for the source 1ES 0229+200 and simulating the pair beam creation rates at various distances in the IGM using the CRpropa Monte Carlo code, we were able to model the beam-plasma interactions in the quasilinear regime and assess their effects on the secondary gamma-ray cascade arrival time at Earth.

Our results show that the instability does not lead to significant angular broadening or time delays sufficient to account for the absence of GeV cascade emission observed in 1ES 0229+200 spectra. The angular spread of the beam, induced by the instability, reaches a steady state where the broadening is confined due to the balance between the instability feedback, the pairs creation and the inverse Compton cooling. The resulting time delay of the GeV-scale cascade emission is only on the order of months, even in the case of flaring activities with higher luminosities.

This limited-time delay is much less than the needed time delay to suppress the secondary GeV emission in the Fermi-LAT observations which has to be at least 15 years corresponding to the minimal possible activity time of the 1ES 0229+200 \citep{Aharonian_2023}. This rules out the pairs' deflection by the instability as a primary mechanism behind the missing cascade emission. Note that higher lifetimes of the 1ES 0229+200 will reinforce this conclusion. Therefore, our findings support the hypothesis that weak intergalactic magnetic fields (IGMF), with strengths greater than femto-Gauss, are more likely responsible for the observed blazer spectra, as they can induce sufficient beam deflection suppressing the GeV-scale secondary emission reaching Earth with the primary emission.

In this study, we didn't consider the Landau damping of the plasma oscillations caused by the MeV-band cosmic-ray electrons \citep{Yang_2024}. The potential impact of this is the damping of the electric fields oblique to the beam direction, resulting in a reduced beam expansion and therefore lower cascade time delays. However, the reduced beam expansion might allow for efficient growth of the parallel plasma oscillation modes that drain the beam energy rather than expanding it. A quantitative estimate, that goes beyond the scope of this paper, is needed to study the impact of such a damping. 


\section*{Acknowledgement}

We thank Andrii Neronov for the useful discussions and insights and Chengchao Yuan for the comments that improved the manuscript. We thank Lorentz Center@Snellius and the organizers of the Primordial Magnetic Fields Workshop for fostering an open and interactive atmosphere. The stimulating discussions and interactions during the workshop were instrumental in shaping the initial ideas of this paper.




\bibliographystyle{aasjournal}
\bibliography{Main} 

\end{document}